\documentclass[conference,compsoc]{IEEEtran}
\usepackage{graphicx}
\graphicspath{{figures/}}
\usepackage{algorithm}
\usepackage{algorithmic}

\usepackage{subcaption}
\usepackage{todonotes}
\usepackage{changebar}
\usepackage{tabularx}

\ifCLASSOPTIONcompsoc
  \usepackage[nocompress]{cite}
\else
  \usepackage{cite}
\fi

\makeatletter
\def\ps@IEEEtitlepagestyle{%
	\def\@oddfoot{\mycopyrightnotice}%
	\def\@evenfoot{}%
}
\def\mycopyrightnotice{%
	{\footnotesize 
		\begin{minipage}{\textwidth}
			\centering
			Copyright~\copyright~2016 IEEE. Personal use of this material is permitted. Permission from IEEE must be obtained for all other uses, in any current or future media, including reprinting/republishing this material for advertising or promotional purposes, creating new collective works, for resale or redistribution to servers or lists, or reuse of any copyrighted component of this work in other works.
		\end{minipage}
		.\hfill}
	\gdef\mycopyrightnotice{}
}

\begin{document}
%
\title{An Efficient Load Balancing Method for Tree Algorithms}

\author{\IEEEauthorblockN{Osama Talaat Ibrahim and Ahmed El-Mahdy}
\IEEEauthorblockA{Computer Science and Engineering Department\\Egypt-Japan University of Science and Technology\\Alexandria, Egypt\\Email: \{osama.ibrahim,ahmed.elmahdy@ejust.edu.eg\}}
}

\maketitle

\begin{abstract}
	Nowadays, multiprocessing is mainstream with exponentially increasing number of processors. Load balancing is, therefore, a critical operation for the efficient execution of parallel algorithms. In this paper we consider the fundamental class of tree-based algorithms that are notoriously irregular, and hard to load-balance with existing static techniques. We propose a hybrid load balancing method using the utility of statistical random sampling in estimating the tree depth and node count distributions to uniformly partition an input tree. To conduct an initial performance study, we implemented the method on an Intel\textsuperscript{\textregistered} Xeon Phi\textsuperscript{TM} accelerator system. We considered the tree traversal operation on both regular and irregular unbalanced trees manifested by Fibonacci and unbalanced (biased) randomly generated trees, respectively. The results show scalable performance for up to the 60 physical processors of the accelerator, as well as an extrapolated 128 processors case.
\end{abstract}


%
\IEEEpeerreviewmaketitle
\section{Introduction}
	According to Moore's law the number of transistors per chip increases with an exponential rate. Multicore design is now becoming widely popular, exploiting the scaling of transistors and overcoming the mainly power constrained scalability for the single-core design; the {ITRS} Roadmap projects that by the year 2022, there will be chips with an excess of 100$\times$ more cores than current multicore processors~\cite{Blake2009, Semiconductors2007}.
	
	This paper investigates the problem of load balancing tree workloads. In general, load balancing is essential for scaling application on multicore systems. It aims to assign workload units to individual cores as uniformly as possible, thereby achieving maximum utilization of the whole system~\cite{xu1997load}.
	
	Tree workloads have special significance in importance as well as in complexity of the load balancing operation owing to their highly irregular structure. From the importance point of view, trees are fundamental in many combinatorial algorithms such as those used in sorting, searching, and optimization (e.g. divide-and-conquer) applications~\cite{Knuth1997, Olivier2007}. The problem facing many tree-based algorithms is that the produced tree is usually unbalanced with respect to the data distribution, having a random number of children per nodes. Thus, it is difficult to statically partition; hence, dynamic load balancing is generally used instead, adding runtime overheads. Therefore, tasks distribution over a group of processors or computers in the parallel processing model is not a straightforward process.
	
	This paper introduces a novel method that combines a one-time quick tree analysis based on random sampling, followed by static partitioning. The method relies on mapping the tree into a linear interval, and subtrees into sub-intervals. Dividing the interval into equal sub-intervals, the method conducts random traversal of corresponding subtrees, estimating the amount of work required for each subtree. The obtained mapping provides for an approximate workload distribution over the linear domain; load-balancing simply reduces to inverse mapping the workload distribution function to obtain corresponding sub-intervals, and hence subtrees; the method further considers adaptive dividing of the considered sub-intervals to account of irregularities on the workload distribution, thereby decreasing sampling error.

	An initial experimental study is conducted on an Intel\textsuperscript{\textregistered} Xeon Phi\textsuperscript{TM} accelerator; we considered two main trees: random and Fibonacci; the former represents irregular unbalanced trees, while the latter represents regular unbalanced trees. Results show better scalability than trivial partitioning of tree with a relative speedup reaching 2$\times$ for 60 cores with projected further growth with increasing number of processors. 
		
	The rest of this paper is organized as follows: Section~\ref{related-work} discusses related work; Section~\ref{methodology} introduces the proposed load-balancing method; Section~\ref{evaluation} provides the experimental study; and finally Section~\ref{conclusions} concludes the paper and discusses future work.
	
	\section{Related Work}
	\label{related-work}
	The load-balancing problem has been addressed previously but not in the manner proposed in this work. Several load balancing algorithms are available, for example Round Robin and Randomized Algorithms, Central Manager Algorithm and Threshold Algorithm. However, these algorithms depend on static load balancing. It requires that the workload is initially known to the balancing algorithm that runs before any real computation.
	
	Dynamic algorithms, such as the Central Queue algorithm and the Local Queue algorithm~\cite{Chakrabarti1994}, introduce runtime overhead, as they provide general tasking distribution mechanisms that do not exploit tree  aspects.
	
	Gursoy suggested several data composition schemes \cite{gursoy2003}; however, they are concerned mainly with tree-based k-means clustering, which does not target general trees.
	
	Another related work is that of El-Mahdy and ElShishiny~\cite{El-Mahdy2011}; that method is the closest to our work in terms of the adopted hybrid static and dynamic approach; however, the method targets the statically structured objects of images, and does not access irregular data structures, such as trees.
	

\section{Suggested Method}
\label{methodology}
Let $ p $ be the number of available processors for which we are partitioning a binary unbalanced tree, as an example without lose of generality. It is worth noting that the tree does not need a data structure, it can also represent control ones, as in recursive branch-and-bound optimization applications.

Our method has three main steps:
\begin{enumerate}
	\item Random unbiased depth probing to estimate the corresponding work for a subtree;
	\item Mapping the measured subtrees work into a one-dimensional linear spatial domain (a scalar); this facilitates the inverse-mapping of the estimated workload;
	\item Utilizing adaptive probing to handle nonlinearities in selecting probes locations.
\end{enumerate}

\subsection{Random Unbiased Depth Probing}\label{subsec:randomProbing}
The method starts by trivially dividing the tree into $ p $ subtrees with the purpose of estimating the average depth of each. This can be simply done by going down in the tree till finding a level that contains $ p $ subtrees.

We consider the node count, as a function of depth, 
to represent a measure of the amount of work in each subtree; however, such function can be changed depending on application. To estimate the node count for each subtree, we perform a series of random depth probes to compute its average depth. Each probe randomly traverses the subtree from its root till hitting a leaf (terminating on a null child) and calculates the path length.

\begin{figure}
	\centering
	\includegraphics[width=\columnwidth]{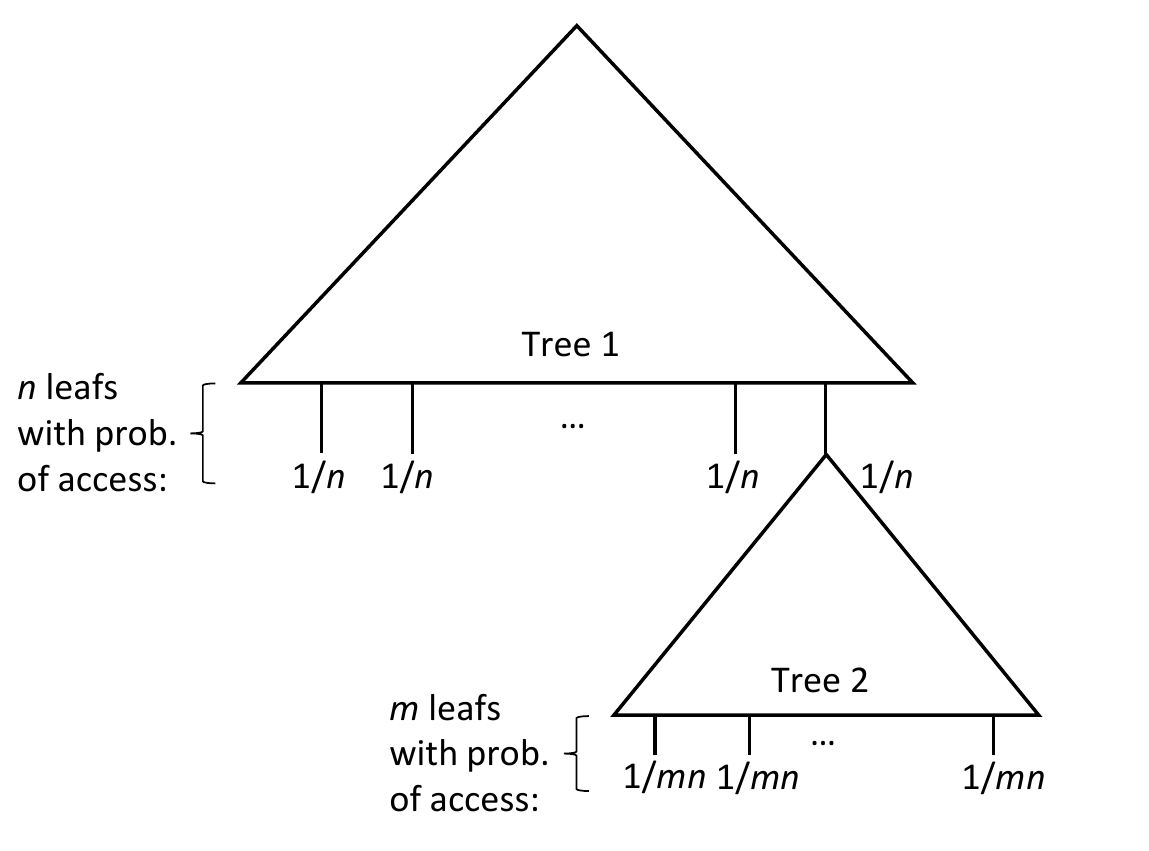}
	\caption{\label{figBiasedSampling}Biased sampling}
\end{figure}

A key issue with such sampling of leaves is its bias; the leaves that occur at shallow depths have much higher chances to be sampled than deeper ones. Fig.~\ref{figBiasedSampling} shows two complete subtrees: Tree~1 and 2; the root of Tree~2 is joined on one of the leaves of Tree~1; the trees have $ n $ and $ m $ leaves, respectively. A random probe will visit a Tree~1's leaf with probability $ 1/n $; whereas it will visit a Tree~2's leaf with probability $ 1/n \times 1/m $, resulting in biased sampling.

To resolve this issue, the obtained depths should be associated with weights; for two leaves separated by delta height of $ h $, the upper leaf would have $ 2^h $ more chance of being probed than the lower one. We, therefore, define a corresponding weight to normalize such effect. The weight $ w_i $ for a depth $ d $ in a probe $ i $ is given by $ w_i = w(d_i) = 2^{d_i} $. Thus, at any probe, $ i $, the weighted average for the $ i $ probes is given by:
\begin{equation}
	avg_i=\frac{\sum_{k=1}^{i}d_kw_k}{\sum_{k=1}^{i}w_k}
	\label{eqAVG}
\end{equation}

Algorithm \ref{algRandomProbe} shows the process of computing the average node count of subtree. The subtree depth is calculated using the above formula for a series of $ i $ probes (lines \ref{lineA}:\ref{lineB}). The corresponding node count is estimated as a function of depth. The running average count is effectively computed at each iteration to make the method more efficient (line \ref{lineAvgDepth}). A sliding window of the last $ N $ counts (line \ref{lineQ}) terminates probing based on a suitable probing stopping criteria ($ \mathit{psc} $) to be less than some threshold; in our implementation we adopt simple relative difference between the node count maximum and minimum values; however, other measures of variance can be used (such as absolute difference, standard deviation, \dots etc.). A simple node count estimator (exponential relation with depth, see Appendix \ref{secCountEstimator}) is used as a fast way to terminate the algorithm, then a better estimator (Algorithm~\ref{algCountEstimator}) is used for measuring the actual node count, based on Knuth~\cite{knuth1975, Vaisman2015}, when termination is reached. Algorithm~\ref{algRandomProbe} is repeated upon each subtree, preferably in parallel.
 
Algorithm~\ref{algCountEstimator} is invoked at the return line of Algorithm~\ref{algRandomProbe}; it takes the count of probes ending at each depth; The loop (lines 2-4) propagates the count up the levels so that recording the total number of times we visited that level. The next loop (lines 7-9) computes the number of nodes at each level by effectively multiplying the maximum number of nodes in the level by the ratio of visits to that level to the total number of visits (i.e. visits to root node), $c(i)/c(1)$. The loops counts nodes for all levels, and returns.

It is worth noting that the depth based estimator has a time complexity of $O ( n (d  a  + b))$ where $n$ is the number of probes, $d$ depth, $a$ is the cost for one level traversal, and $b$ is arithmetic operation cost. This is due to performing only one arithmetic operation per probe. Whereas using solely the  Knuth-based estimator requires $O ( n  d (a + b) )$ computations as every probe requires $d$ arithmetic operations for each level traversal. Generally the cost of traversal is less than arithmetic operations, hence $a << b$ and thus our algorithm is approximately $O ( n )$, whereas the other is $O ( n  d)$; however, our implementation adds memory complexity of $O ( d )$ to store records where as the slow estimator requires $O( 1 )$.

\begin{algorithm}
	\begin{algorithmic}[1]
		\REQUIRE Subtree, $ \mathit{psc} $
		\STATE $ \mathrm{current}\gets $ subtree root
		\STATE $ \mathrm{sum} \gets 0 \texttt{	//Eq.~\ref{eqAVG} numerator} $ 
		\STATE $ \mathrm{num} \gets 0 \texttt{	//Eq.~\ref{eqAVG} denominator} $
		\STATE $\mathrm{avgQ} \gets$ FIFO queue of length $ N $, initialized to zeros \label{lineQ}
		\STATE $ \mathrm{depthCounter} \gets \{0\} $ \texttt{//Dynamic array}
		\REPEAT \label{lineA}
		\STATE $ d \gets 0 $
		\WHILE{$\mathrm{hasChild(current)}$}
		\STATE \texttt{//visit right or left child at} \STATE \texttt{//random}
		\IF{$ \mathrm{randBoolean()=0} $}
		\STATE $ \mathrm{current} \gets \mathrm{LeftChild(current)} $
		\ELSE
		\STATE  $ \mathrm{current} \gets \mathrm{RightChild(current)} $
		\ENDIF
		\STATE $ d++ $
		\ENDWHILE
		\STATE $ \mathrm{depthCounter}(d)++$
		\STATE $ \mathrm{sum} \gets \mathrm{sum} +  d \times 2^d$
		\STATE $ \mathrm{num} \gets \mathrm{num} +  2^d $
		\STATE $ \mathrm{avgDepth} = \mathrm{sum/num} $ \label{lineAvgDepth}
		\STATE $ \mathrm{avgQ.add(\mathrm{FastNodeCount(avgDepth)})} $
		\UNTIL $\frac{\max(avgQ)-\min(avgQ)}{\max(avgQ)} < psc $\label{lineB}
		\RETURN $ \mathrm{NodeCount(depthCounters)} $
	\end{algorithmic}
	\caption{Estimating the average node count of a subtree}
	\label{algRandomProbe}
\end{algorithm}
\begin{algorithm}
	\begin{algorithmic}[1]
		\REQUIRE $ c $: depth history counts
		\FORALL {$ i \gets \mathrm{length}(c)-1 $ \textbf{to} 1}
		\STATE \texttt{//Accumulate}
		\STATE $ c(i) \gets c(i) +c(i+1) $
		\ENDFOR
		\STATE $ \mathrm{count} \gets 1 $ \texttt{//root}
		\STATE $ \mathrm{level} \gets 1 $
		\FORALL {$ i \gets $ 2 \textbf{to} $ \max(\mathrm{depths}) $ }
		\STATE $ \mathrm{count} \gets \mathrm{count} + \mathrm{level} \times \frac{c(i)}{c(1)}$
		\STATE $ \mathrm{level} \times= 2 $
		\ENDFOR
		\RETURN	$ \mathrm{count} $
	\end{algorithmic}
	\caption{Estimating the node count from depth history}
	\label{algCountEstimator}
\end{algorithm}
\subsection{Subtree Work Mapping}
This step generates a work distribution function by making a linear mapping between the subtrees' work (node count) into a sub-interval on a one-dimensional domain. In other words, any subtree is labeled with an interval.

All the nodes within the tree are effectively covered by the interval [0,1]. It represents the interval of the root node. The children of a node divide its interval into equal sub-divisions, each representing the corresponding node's subtree regardless of the subtrees sizes. Thus, if there are $ m $ nodes in a level, each node $ i $ has the sub-interval:$ [(i-1)/m, i/m] $.

\begin{figure}
	\centering
	\includegraphics[width=\columnwidth]{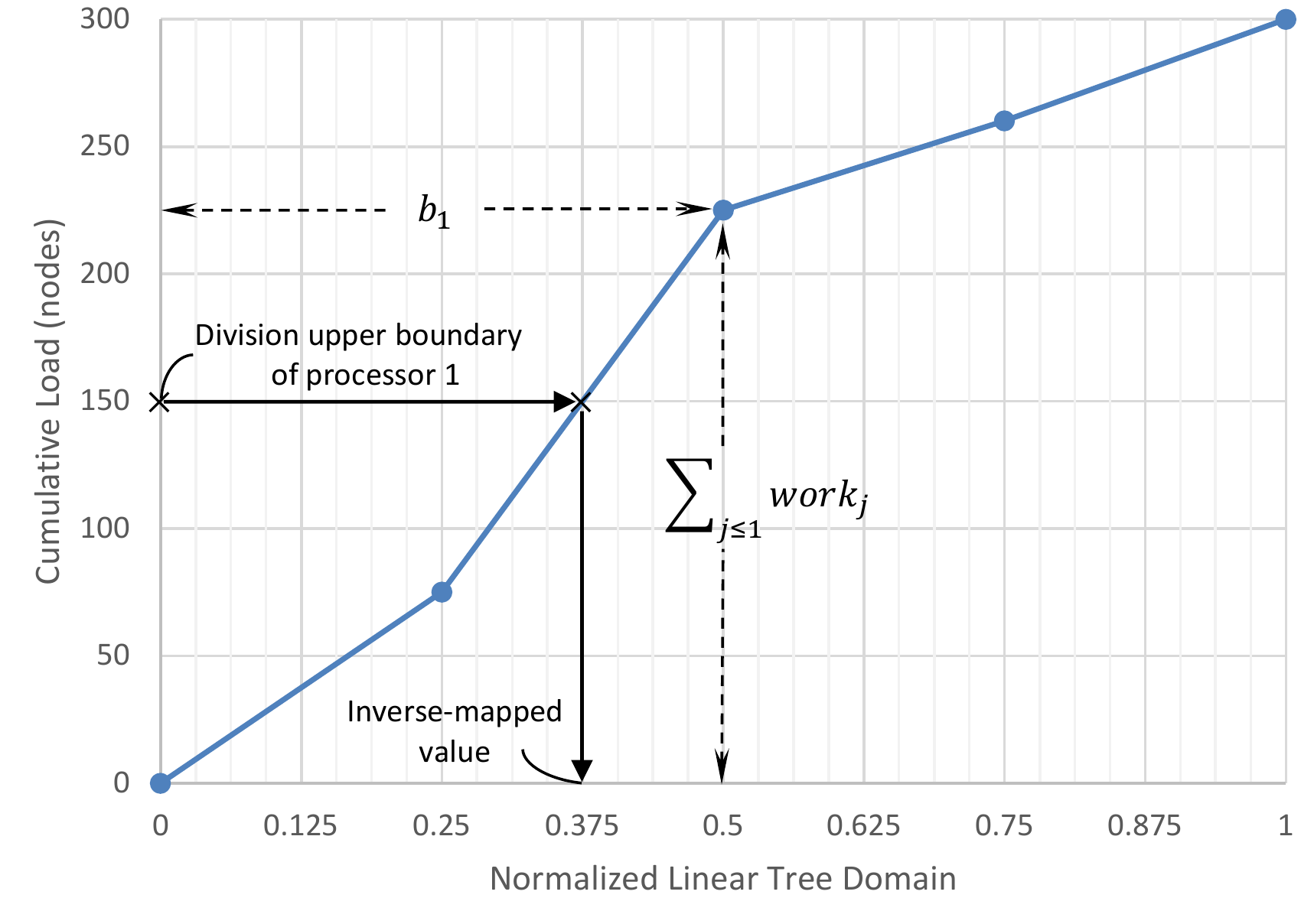}
	\caption{\label{figWorkDistribution}Example of work distribution and inverse mapping}
\end{figure}

The work estimate (node count) obtained from the previous step is accumulated and associated with the corresponding interval's upper bound. The accumulated node count for each subtree is equal to the sum of its count and its previous subtrees counts. Thus, if the work for subtree $ i $ is $ work_i $ and the corresponding interval is $ [a_i, b_i] $, then we define the mapping:  $ b_i \rightarrow \sum_{j \leq i}\mathit{work}_j  $; where $ b_i \in [0,1]$  ($x$-axis), and $\sum_{j \leq i} \mathit{work}_j$ is the corresponding cumulative work for the interval ending with $b_{i}$; all other non-boundary points are piece-wise linearly interpolated. The curve
is expected to be monotonically increasing. A mapping example with hypothetical values is illustrated in Fig.~\ref{figWorkDistribution}, The $ x $-axis represents the linear space values for the whole tree (i.e. the interval $ [0,1] $), and the $ y $-axis represents the average accumulated weighted depth for $ 4 $ subtrees ($ p=4 $).

Due to the accumulation process, the maximum value in the $ y $-axis range is  the total estimated work value, which we seek to partition over the available processors. Thus, we divide this range into $ p $ equal divisions, representing the corresponding load. Then, we inverse-map the divisions to obtain the intervals that represent the work load for each processor, and hence the corresponding set of subtrees. The inverse-mapping processes over the plot can be done mathematically using straight-line equation. We use the straight-line part of curve where each division boundary intersects. In Fig.~\ref{figWorkDistribution}, the total work value is 300, which will be divided into $ p=4 $ divisions; then, the optimal work division for processor 1, for example, is $ 75:150 $, whose upper value is inverse-mapped through the curve to result in the $ 0.375 $ at the tree linear domain. 

In general, a processor inverse-mapped sub-interval covers many subtrees. Algorithm~\ref{algFinalDivid} shows the steps for finding and pruning off all the required subtrees for an interval into a result set. The algorithm starts at the node identified by the processor interval's upper bound; it would generally be a left child. The algorithm clips this node defining a new subtree in the result set (lines \ref{line:Clipping}, \ref{line:union}). Its parent extends the coverage range beyond the sought target sub-interval; so the algorithm then traverses up the tree until reaching the first right child node; it then takes its left sibling as a new subtree in the result set. The algorithm recursively repeats until reaching the root, then terminates.

\begin{algorithm}
	\begin{algorithmic}[1]
		\REQUIRE $ \mathrm{root} $, current processor interval's end
		\STATE $\mathrm{resultSet} \gets \phi$ \texttt{	//empty array of subtrees}
		\STATE $ \mathrm{current} \gets \mathrm{Node(intervalEnd)} $
		\WHILE{$ \mathrm{current} \ne  \mathrm{root} $}
		\IF{$ \mathrm{IsLeftChild(current)} $}
		\STATE $ \mathrm{Tree(root)} = \mathrm{Tree(root)} - \mathrm{Tree(current)} $ \label{line:Clipping}
		\STATE $ \mathrm{resultSet} = \mathrm{resultSet} \cup \mathrm{Tree(current)} $\label{line:union}
		\REPEAT
		\STATE \texttt{//Go up till hitting the root or}
		\STATE \texttt{//a right child}
		\STATE $ \mathrm{current} = \mathrm{Parent(current)} $
		\UNTIL {$ \mathrm{current} = \mathrm{root} \lor \mathrm{IsRightChid(current)} $} 
		\ELSIF{$ \mathrm{IsRightChid(current)} $}
		\STATE $ \mathrm{current} \gets \mathrm{LeftSibling(current)} $
		\ENDIF
		\ENDWHILE
		\RETURN $ \mathrm{resultSet} $
	\end{algorithmic}
	\caption{Finding the final workload of a processor}
	\label{algFinalDivid}
\end{algorithm}
The algorithm is applied once for each $ p-1 $ processors to generate the corresponding result sets; for the last processor, the result set trivially contains one subtree which is the remaining part of the tree starting from the root. For a proper division, the order of processors should maintain, from processor $ 0 $ to the last one; the pruning is based only on the interval's end, it considers $ 0 $ or previous processor interval's end as the start of the current one's interval.

\subsection{Adaptive Probing}

For the sake of accuracy in the mapping process and because the work distribution is approximated by few points only, the division boundary, which is a depth value itself, should be so close to an original point of the work distribution. The closer to that original point, the more accurate we expect the results to be. However, generally, we cannot guarantee such condition, therefore adaptive probing mends that by dynamically creating another probe point on the work distribution, which decreases the approximation error between the fitted points and the actual workload. We compare the work difference between the division boundary and the closest point to be a factor of the optimal division, which is the total work value divided by $ p $. We name this factor as the adaptive stopping criteria ($ asc $).

Algorithm~\ref{algAdaptiveProbing} illustrates the adaptive probing process. Each inverse-mapping iteration targets a straight line in the workload distribution (line \ref{line:IntersectionLine}). At the interval label of its middle (lines \ref{line:xMean}, \ref{line:probeStart}), the adaptive probing re-probes once more if the adaptive criteria is not satisfied. This results in another node count value in the work distribution, which represents the subtree of the current node's left child. The algorithm iteratively checks again the adaptive criteria and re-probes another point till being satisfied. Generally, we specify the adaptive stopping criteria as a percentage of the current processor node count workload, which was previously calculated by dividing the $ y $-range into $ p $ workloads (line \ref{line:asc}).
\begin{algorithm}
	\begin{algorithmic}[1]
		\REQUIRE Tree, $ \mathrm{WD} $: workload distribution, $ y $:  current processor's division boundary
		\STATE \label{line:IntersectionLine} $ \overline{P_1P_2} \gets \overline{(x_1,y_1)(x_2,y_2)}: y_1<y<y_2 \land P_1, P_2 \in \mathrm{WD}$    
		\STATE $ \mathrm{probStart} =  \mathrm{LeftChild(IntervalToNode}(x_1, x_2)) $ \label{line:probeStart}
		\WHILE{$ \min(y-y_1,y_2-y) \leq \mathit{asc} \frac{y_{\max}}{p} $} \label{line:asc}
		\STATE $ P_{\mathit{new}} \gets [x_\mathit{new}, y_\mathit{new}]$ \texttt{	//new empty point}
		\STATE $ x_\mathit{new} \gets \frac{x_1+x_2}{2} $ \label{line:xMean}
		\STATE $ y_\mathit{new} \gets \mathrm{RandomProbing(probStart)} $
		\STATE  $ y_\mathit{new} = y_\mathit{new}+y_1 $ \texttt{		//Accumulate}
		\IF{$ y_1<y<y_\mathit{new} $}
		\STATE $ P_2 \gets P_\mathit{new} $
		\STATE \texttt{//for next reprobeing, if required}
		\STATE $ \mathrm{probStart} \gets \mathrm{LeftChild(probStart)}$
		\ELSIF{$ y_\mathit{new}<y<y_2 $}
		\STATE $ P_1 \gets P_\mathit{new} $
		\STATE $ \mathrm{probStart} \gets \mathrm{LeftChild(RightSibling(probStart))} $
		\ENDIF
		\ENDWHILE
		\RETURN $ \overline{P_1P_2} $ \texttt{	//to  use in inverse mapping}
	\end{algorithmic}
	\caption{Adaptive probing}
	\label{algAdaptiveProbing}
\end{algorithm}
\section{Experimental Evaluation}
\label{evaluation}
\subsection{Experimental Environment}
The suggested algorithm has been implemented using C++ for a tree traversal benchmark based on the well-known Pthreads programming model. It is executed on an Intel\textsuperscript{\textregistered} Xeon Phi\textsuperscript{TM} 5110P accelerator with 60 cores running at 1.053 GHz~\cite{xeonPhi}, allowing for studying the scalability of the method.

For this initial study, we choose tree traversal as it is a fundamental operation in many important applications such as search and in counting problems \cite{blanchet2009}. We also considered two unbalanced input trees: a regular and highly irregular. For the former, we consider the well-known Fibonacci tree containing 2.7 million nodes.

For the latter, we considered generating unbalanced random trees. Directly inserting random nodes results in generally balanced trees; we thus provide some bias in the generation of the tree. In particular, we generate a list of sorted numbers, then swap the locations of random pairs; the number of swapping pairs is set to 50\% of the tree size, so theoretically 100\% of elements are randomly swapped. The tree is then constructed by inserting the numbers in an empty binary search tree. The tree has 1 million nodes.

\subsection{Experimental Results \& Discussion}
\subsubsection{Fibonacci Tree}
\begin{figure}
	\begin{subfigure}{\columnwidth}
		\centering
		\includegraphics[width=\columnwidth]{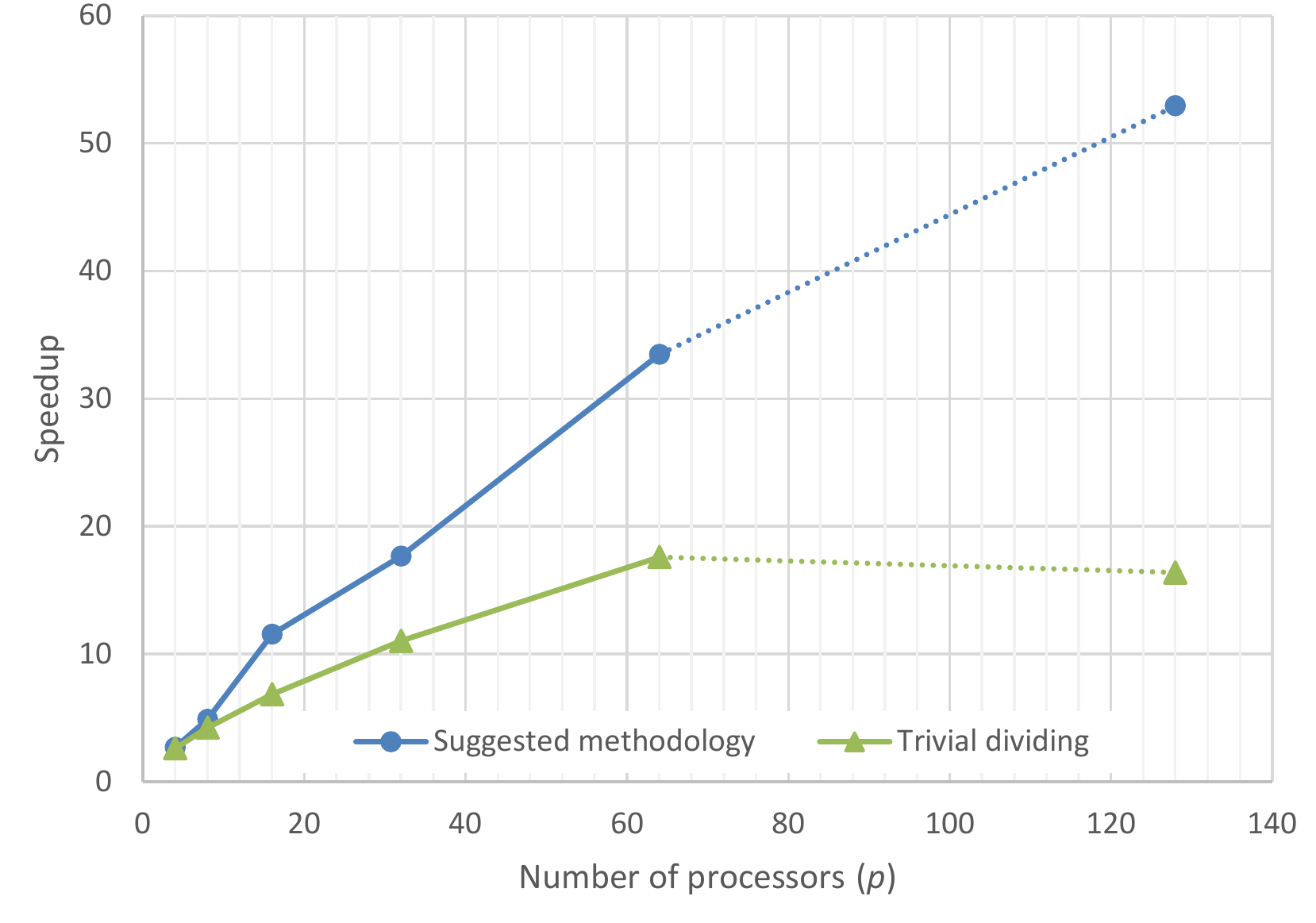}
		\caption{Speedup for Fibonacci Tree Traversal}
		\label{figSpeedupFibonacci}
	\end{subfigure}
	\begin{subfigure}{\columnwidth}
		\centering
		\includegraphics[width=\columnwidth]{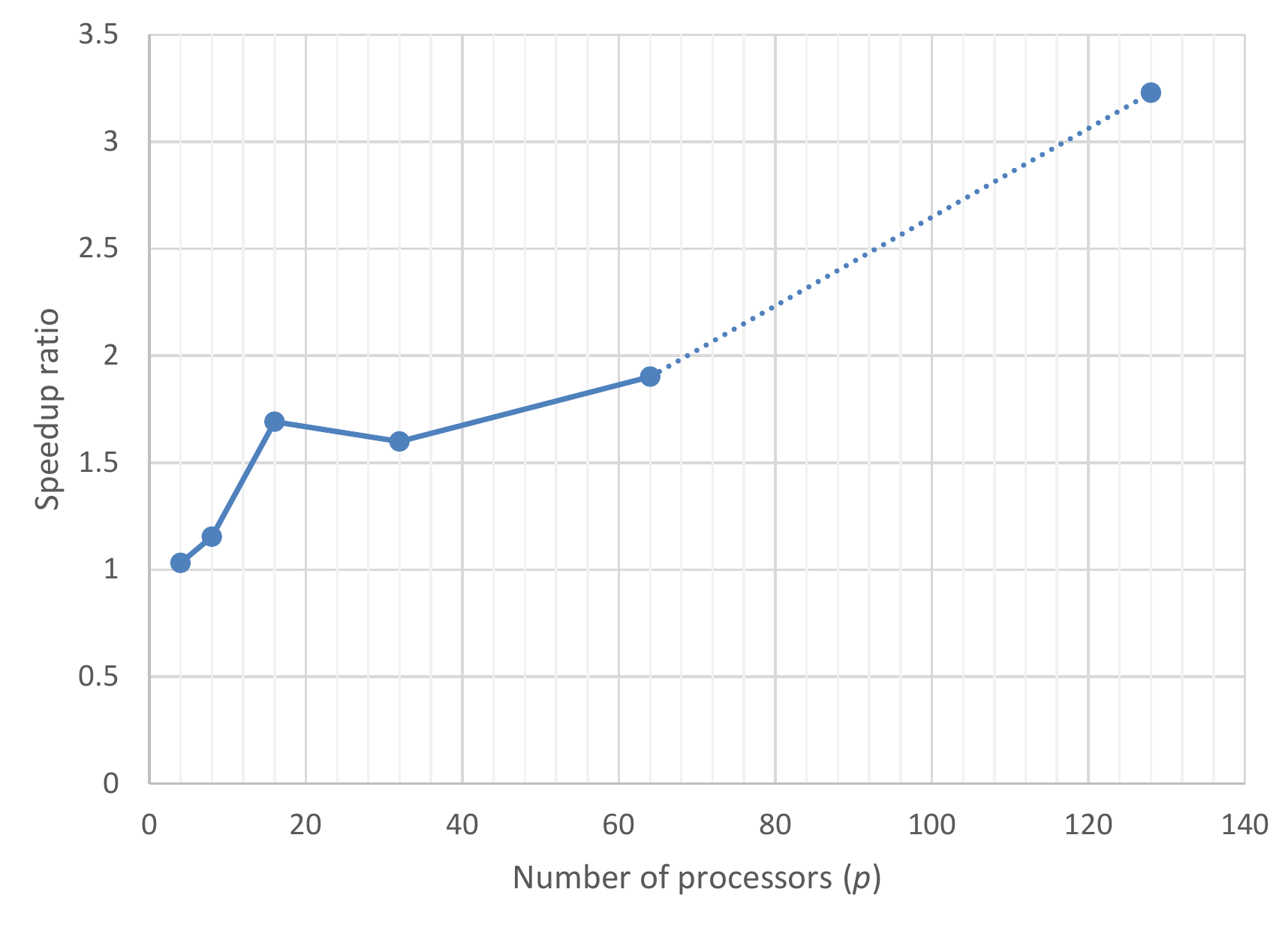}
		\caption{Speedup Over Trivial Dividing for Fibonacci Tree Traversal}
		\label{figSpeedupRatioFibonacci}
	\end{subfigure}
	\caption{Fibonacci Tree Traversal Results}
\end{figure}
Fig.~\ref{figSpeedupFibonacci} represents the speedup due to the suggested method for increasing number of processors for Fibonacci tree traversal. The figure also includes speedup results for the trivial partitioning method (Section~\ref{subsec:randomProbing}). The latter assesses the degree of imbalance of the tree. The results show up to 33.5$\times$ speedup compared to 17.6$\times$ in trivial partitioning. The figure shows a linear behavior with increasing the number of available processors; the speedup is expected to continue going up till reaching 53$\times$ at 128 processors, based on the dotted extrapolation. Fig.~\ref{figSpeedupRatioFibonacci} shows the speedup ratio between the suggested method and the trivial dividing, reaching 1.9$\times$ at 64 processors. We notice that for small number of processors there is no speedup over trivial dividing. This is mainly due to method overheads. The speedup ratio also follows a linear behavior; the dotted extrapolation curve projects that this ratio would increase with larger numbers of processor.

\subsubsection{Random Tree}
\begin{figure}
	\begin{subfigure}{\columnwidth}
		\centering
		\includegraphics[width=\columnwidth]{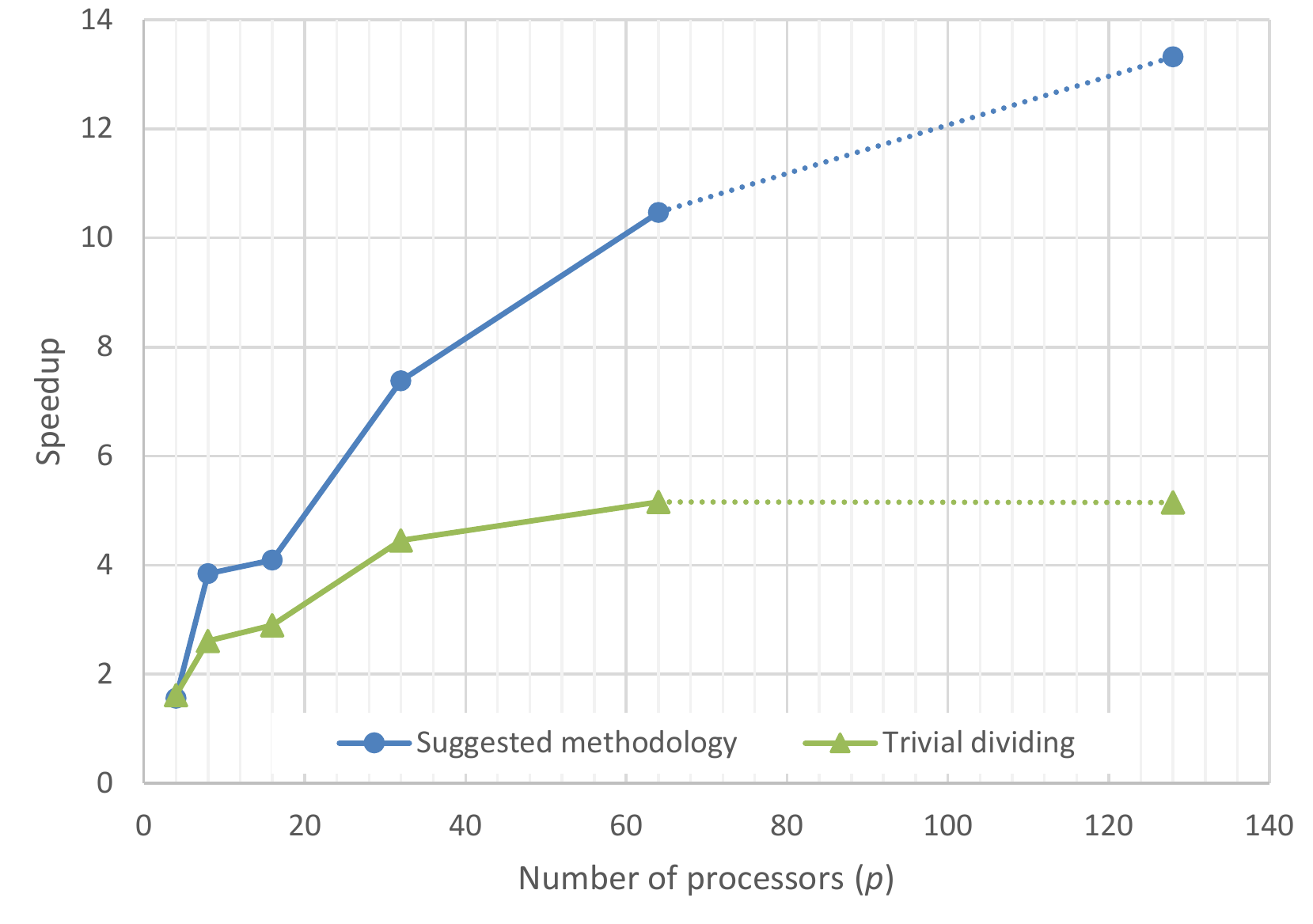}
		\caption{Speedup for Random-Tree Traversal}
	\end{subfigure}
	\begin{subfigure}{\columnwidth}
		\centering
		\includegraphics[width=\columnwidth]{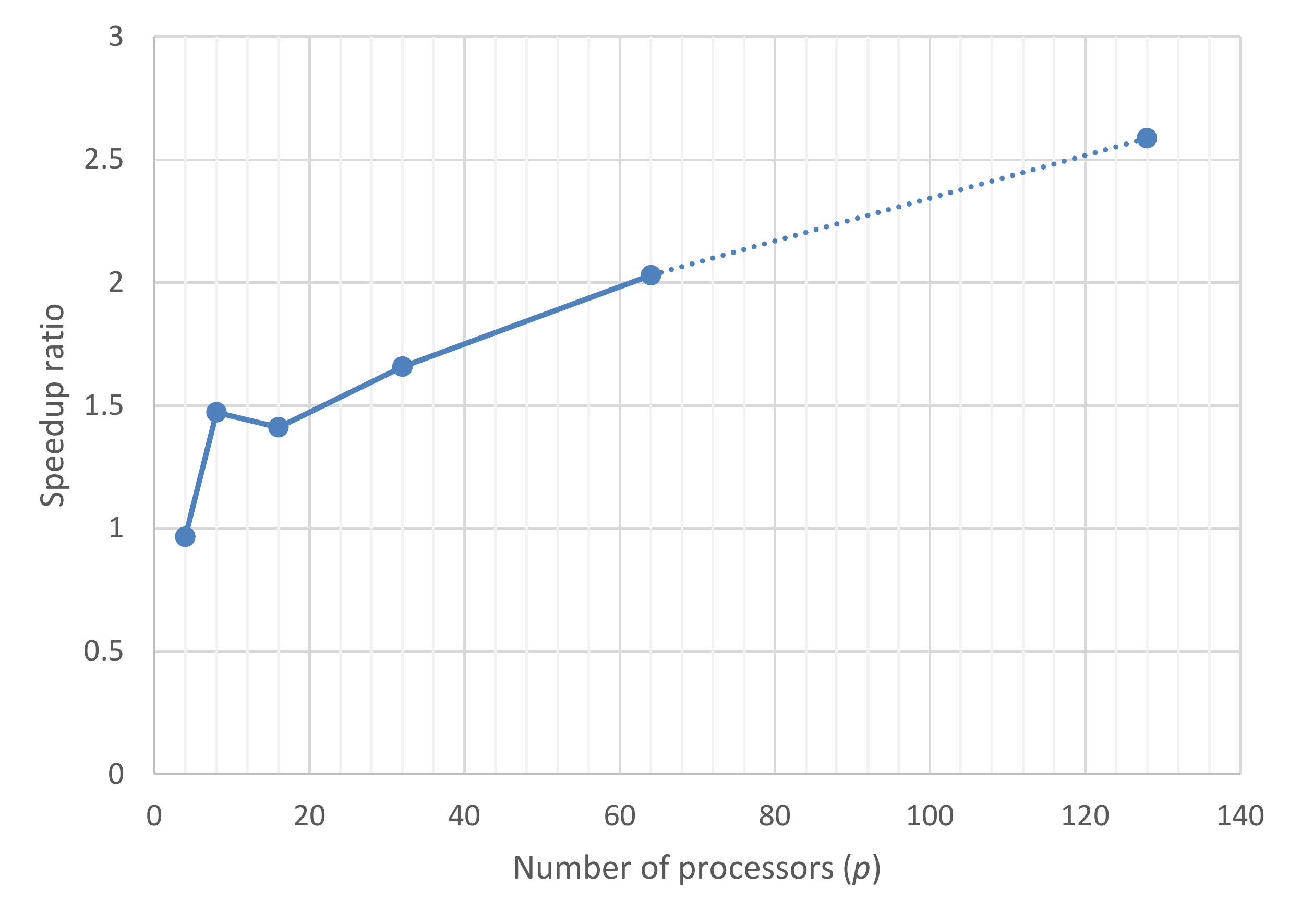}
		\caption{Speedup over Trivial Dividing for Random-Tree Traversal}
	\end{subfigure}
	\caption{Random-Tree Traversal Results}
	\label{figRandomResults}
\end{figure}
Fig.~\ref{figRandomResults} represents the same above results but over the random tree. The results show 10.5$\times$ total speedup for 64 processors, with projected speedup of 13.3$\times$ for 128 processors. The results are lower than the Fibonacci's case, owing to high irregularity of the random tree, it is not likely to find a tree in an application that is completely random. The results might provide a rather pessimistic case results for the general class of tree traversal applications.


\subsubsection{Stopping Criteria}
\begin{figure}
	\begin{subfigure}{\columnwidth}
		\centering
		\includegraphics[width=\columnwidth]{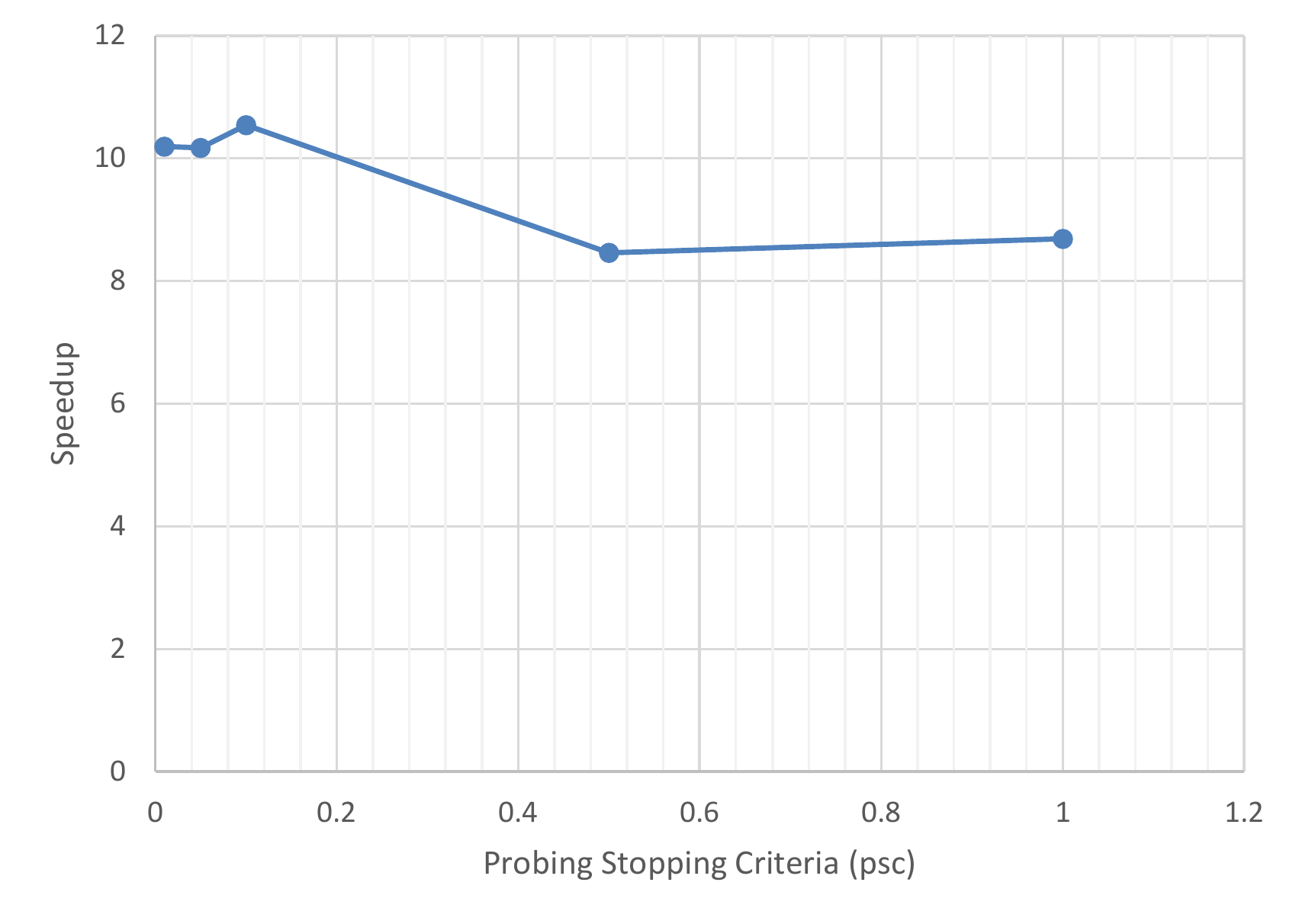}
		\caption{\label{figPSC}The Effect on Speedup Values}
	\end{subfigure}
	\begin{subfigure}{\columnwidth}
		\centering
		\includegraphics[width=\columnwidth]{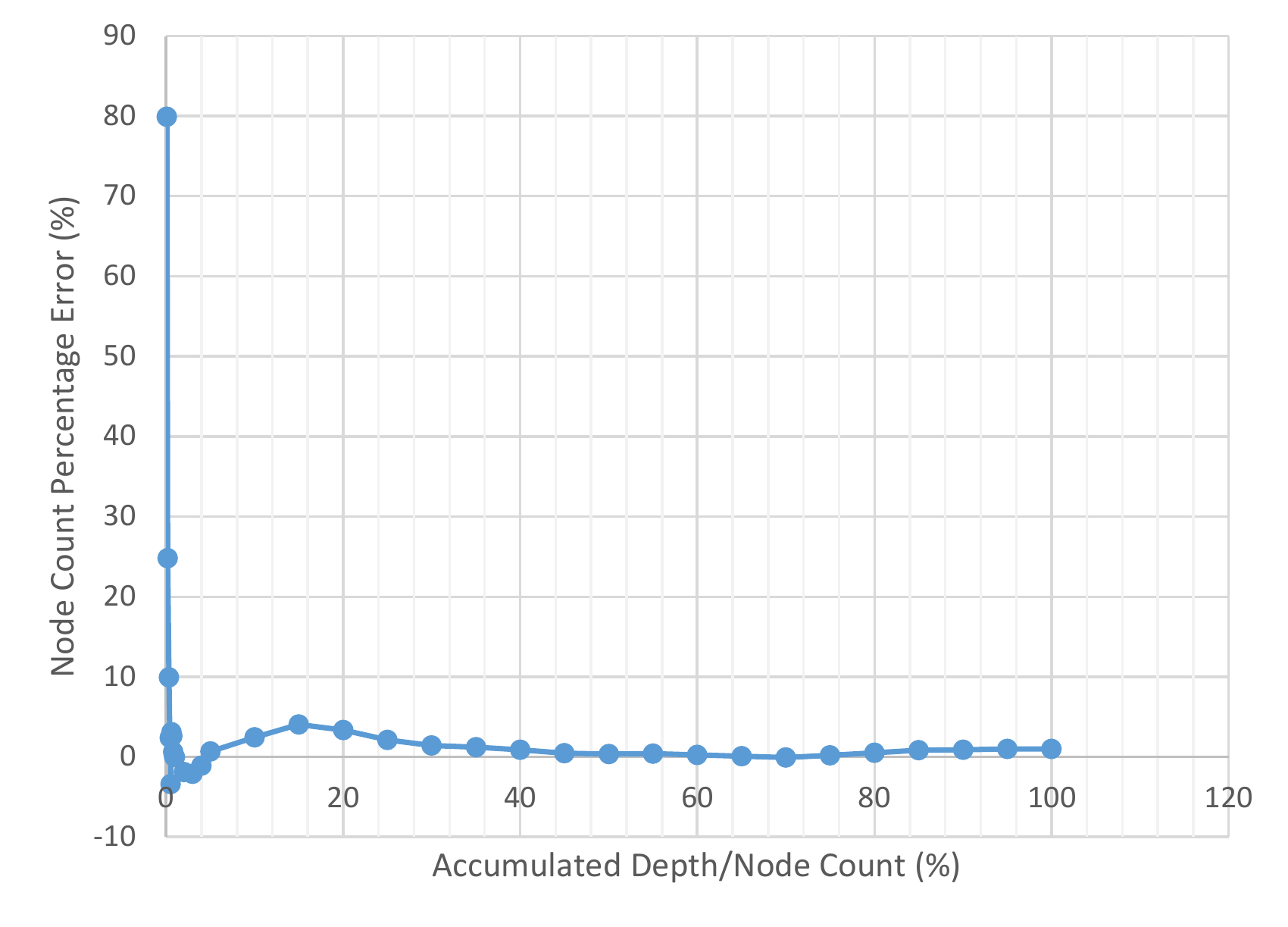}
		\caption{\label{figCountError}The Effect of Accumulated Depth/Node Count}
	\end{subfigure}
	\caption{The Effect of Probing Stopping Criteria $ \mathit{psc} $}
\end{figure}
\begin{figure}
	\begin{subfigure}{\columnwidth}
		\centering
		\includegraphics[width=\columnwidth]{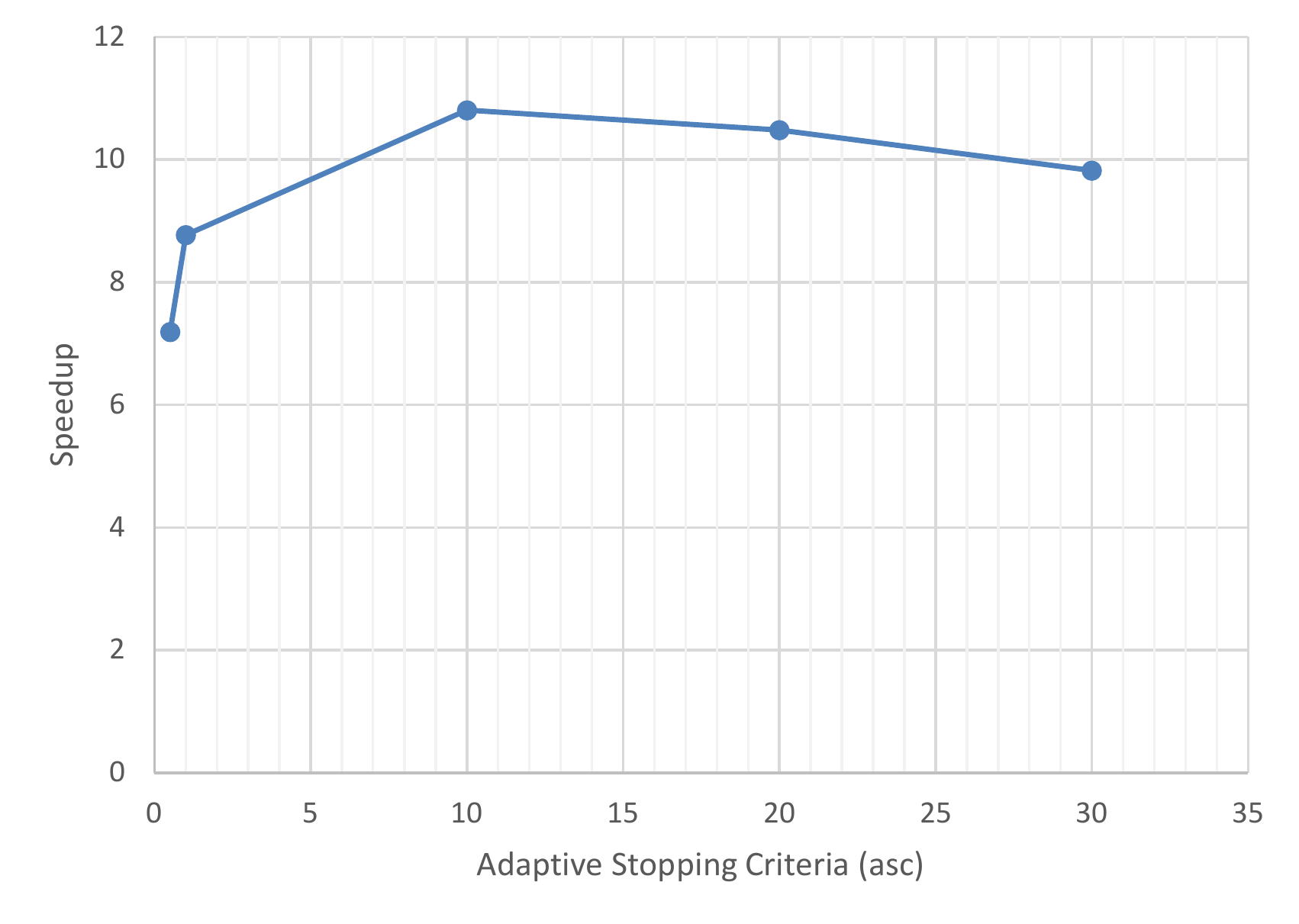}
		\caption{\label{figASC}The Effect on Speedup Values}
	\end{subfigure}
	\begin{subfigure}{\columnwidth}
		\centering
		\includegraphics[width=\columnwidth]{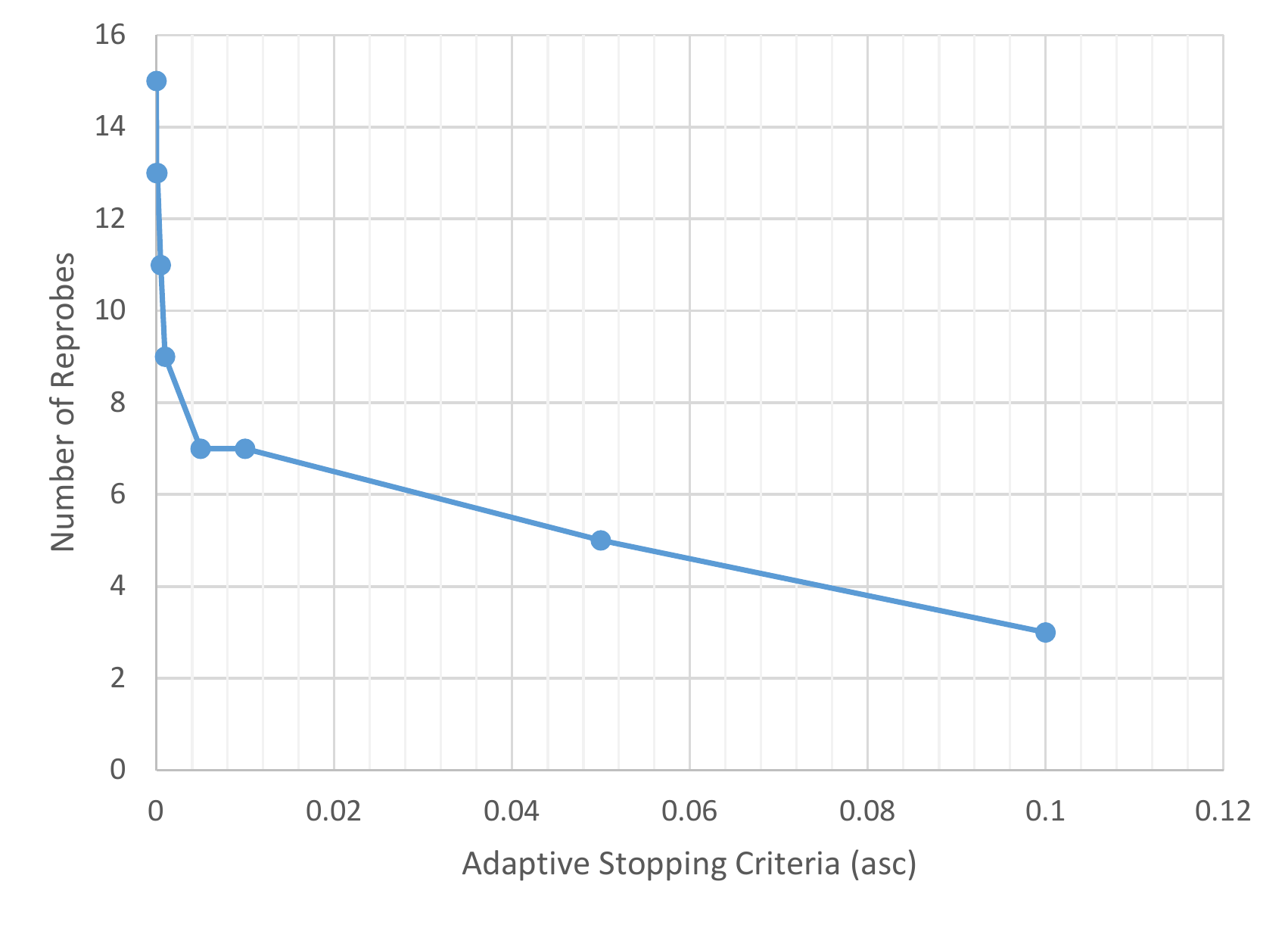}
		\caption{\label{figReprobes}The Effect on Number of Reprobes}
	\end{subfigure}
	\caption{The Effect of Adaptive Stopping Criteria $ \mathit{asc} $}
\end{figure}
There is a trade-off between probing and adaptive stopping criteria on one hand, and the speedup gained on the other hand. More precisely, applying more depth probes enhances the division accuracy, resulting in smaller time values for the traversal itself, but also increases the time required to run the balancing algorithm resulting in higher total running time. The above results are obtained under $ \mathit{psc} = 0.1 $ and $ \mathit{asc} = 10 $. To illustrate the effect of the algorithm criteria on the speedup, we run the benchmark at different values of $ \mathit{psc} $ and $ \mathit{asc} $.

Fig.~\ref{figPSC} shows how changing the probing stopping criteria affects the speedup value at constant $ p=64 $ and $ \mathit{asc}=10 $. With decreasing $ \mathit{psc} $, the speedup is enhanced till reaching a peak point at $ \mathit{psc} = 0.1 $; after which the speedup is reduced. At certain value of $ \mathit{psc} $ there is a balance between the probing time load and the probing accuracy which is translated into speedup gain. Over this value, increasing $ \mathit{psc} $ value leads to less probes performed, hence less probing time with more speedup enhancement, but also less accuracy which means more speedup degradation, and vice versa in case of decreasing $ \mathit{psc} $ which also results in the same overall degradation. This $ \mathit{psc} $ value varies with different applications. 

Another estimation of the effect of the $ \mathit{psc} $ is to consider the percentage of total number of visited nodes for all probes over the actual number of nodes in the tree, against the estimation error. As $ \mathit{psc} $ decreases the number of visited nodes increases. Fig.~\ref{figCountError} shows the results indicating fast convergence when visiting 10\% of nodes, and stability upon reaching 40\%.


Fig.~\ref{figASC} shows the effect of changing the adaptive stopping criteria over speedup at 64 processors and $ \mathit{psc}=0.1 $. It, as well, presents the same above behavior of changing $ \mathit{psc} $. This is due to that changing $ \mathit{asc} $ affects also the same number of probes mentioned above. The effect of $ \mathit{asc} $ on the number of reprobes is shown in Fig.~\ref{figReprobes}.

\subsubsection{Algorithm Efficiency}
\begin{figure}
	\begin{subfigure}{\columnwidth}
		\centering
		\includegraphics[width=\columnwidth]{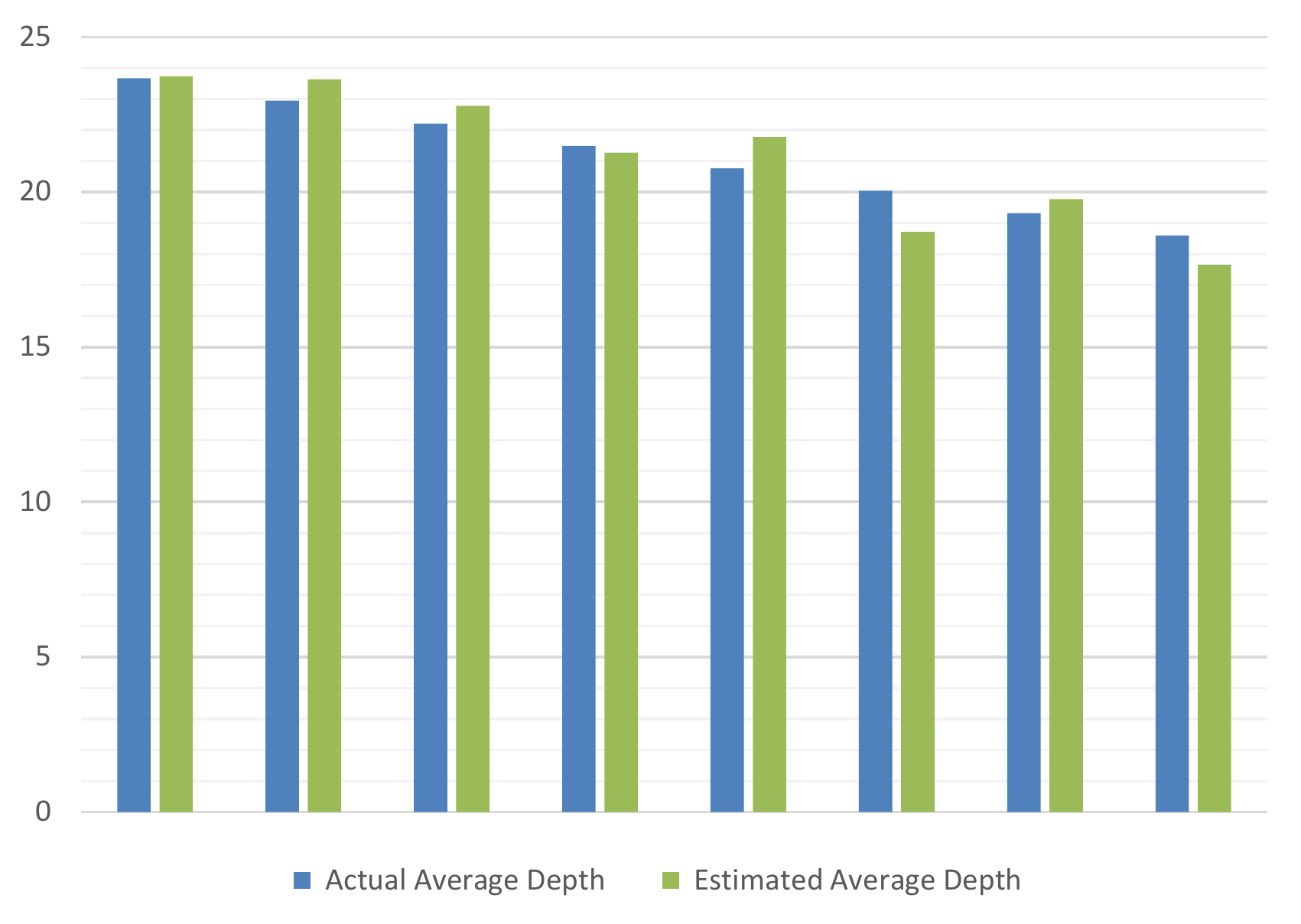}
		\caption{\label{figActualEstimatedDepth}Comparison between Actual \& Estimated Average Depths}
	\end{subfigure}
	\begin{subfigure}{\columnwidth}
		\centering
		\includegraphics[width=\columnwidth]{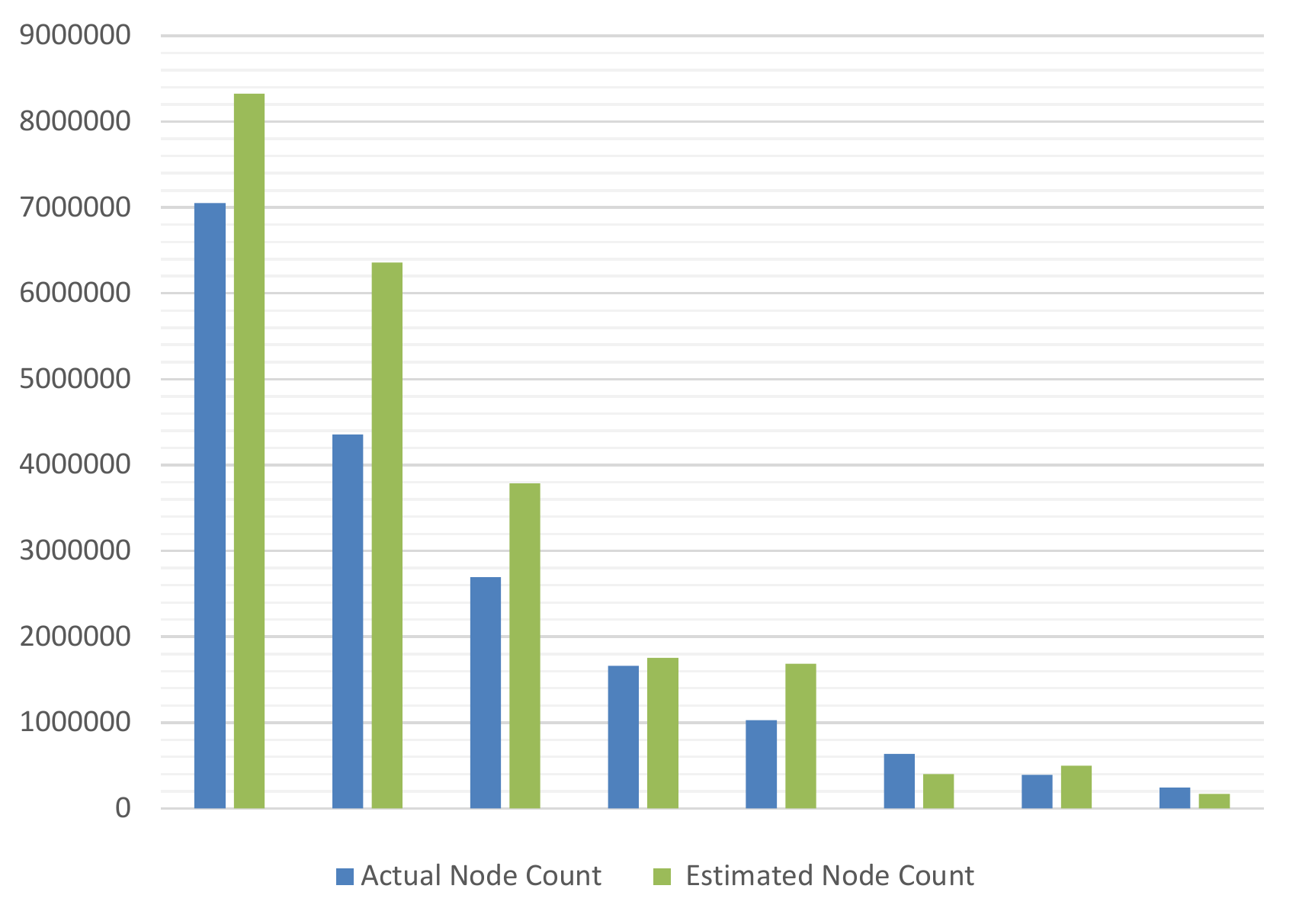}
		\caption{\label{figActualEstimatedCount}Comparison between Actual \& Estimated Average Node Counts}
	\end{subfigure}
	\caption{The Efficiency of Depth and Node Count Estimators}
	\label{figActualEstimated}
\end{figure}
To test the load balance algorithm, three experiments was performed. First and second is to evaluate the efficiency of the depth and node count estimators, the depth and node count calculated values using our estimators should be compared to their actual values. Fig~\ref{figActualEstimated} presents this comparison results. The test was performed to the entire tree at several sizes. Fig.~\ref{figActualEstimatedDepth} considers the random depth estimator and Fig.~\ref{figActualEstimatedCount} concerns about the node count estimator. The latter clearly provides for better accuracy; however, it is worth noting that the former only provides for a stability criteria; one possible scenario it to rely only on the depth, however, that would require substantially increasing the number of probes, as count is not linear with the depth, and hence error is.

\begin{figure}
	\begin{subfigure}{\columnwidth}
		\centering
		\includegraphics[width=\columnwidth]{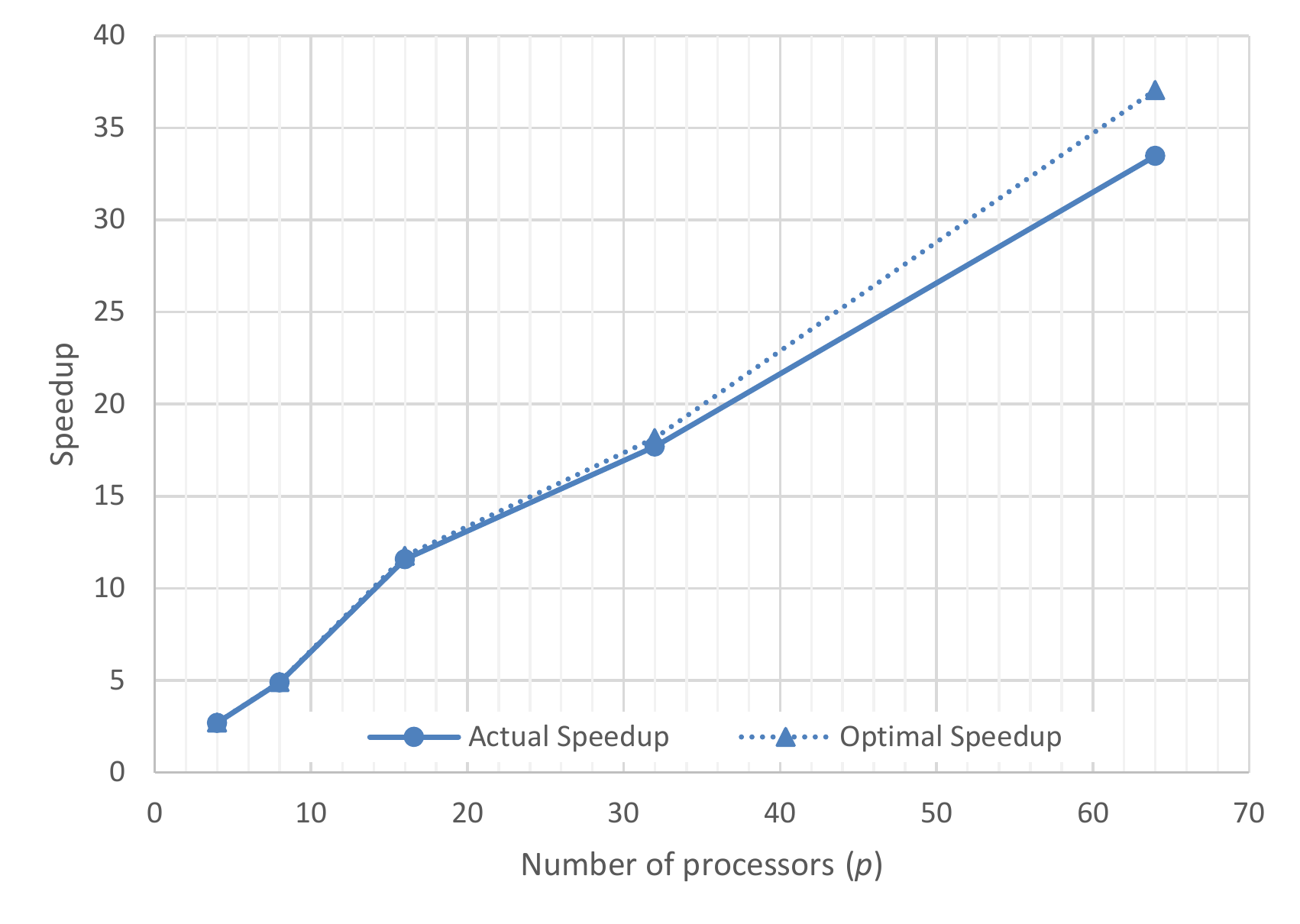}
		\caption{\label{figCountSpeedup}Comparison between Actual and Optimal Speedup}
	\end{subfigure}
	\begin{subfigure}{\columnwidth}
		\centering
		\includegraphics[width=\columnwidth]{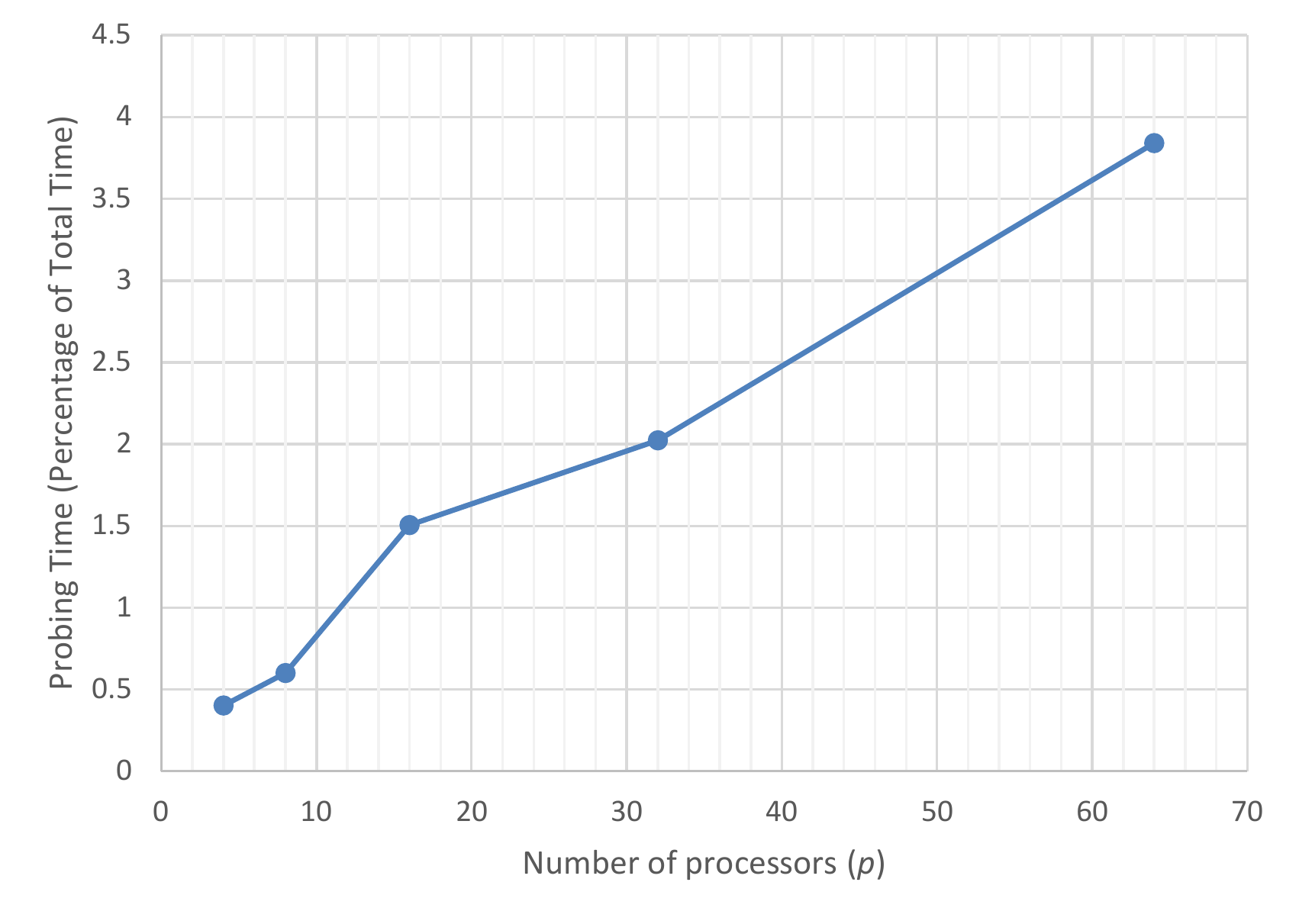}
		\caption{\label{figProbeTime}Probing Overhead}
	\end{subfigure}
	\caption{The Algorithm Runtime Overhead}
\end{figure}
The third experiment is to evaluate the algorithm runtime overhead. This done by two runs: The first is to compare the obtained speedup with its optimal value. The optimal value is calculated supposing that there is no runtime overhead. Furthermore, to eliminate any other circumstances, the optimal speedup value is calculated as the node count speedup, i.e. the ratio between the node count in serial execution to the max node count of all processors in parallel. Fig.~\ref{figCountSpeedup} presents this observation for the above Fibonacci tree. The second one is to find the probing overhead time as a percentage of the total benchmark runtime, this is illustrated in Fig.~\ref{figProbeTime}. The probe time does not exceed 5\% of the total runtime. This percentage is increasing with the number of processors. The reason for this is that probing time reduces much slower than overall speedup, indicating that the convergence is not linear. It is worth-noting that the probing processes should be done in parallel over all the processors to reduce the runtime workload. This is left for future work; for this paper we count the maximum probing time of all processors to be the probe time reported above. Another point to mention is that the probe time includes the time required to run the node count estimator, so it includes all the overheads.
\section{Conclusions and Future Work}
\label{conclusions}
This paper proposes a novel tree load-balancing using statistical random sampling. Our method allows for unbiased sampling of leaves and provides for a linear mapping of estimated tree work into a spatial one-dimensional domain. Furthermore, it utilizes adaptive probing to increase the mapping accuracy and get better results for load values. Results on an Intel\textsuperscript{\textregistered} Xeon Phi\textsuperscript{TM} accelerator shows scalable results for both unbalanced regular Fibonacci and irregular random trees without introducing dynamic scheduling overheads.

This method has the following advantages: 1) It provides fast load balancing for complex tree-based applications, 2) It requires modest memory resources for such process, making it suitable and applicable to even modest embedded devices, and 3) The method achieves significant scalable speedup with increasing the number of processors.

Future work include extensive evaluation of the proposed method over many other typical tree applications, including recursive applications. Further work is also needed to access the effect of the quality of the random number generator on the results. Moreover, the effect of communication is not considered, and it is an important topic for future work. Finally, the conversion behaviour for the chosen estimators requires further investigation for further scalability.

\section*{Acknowledgments}
The first author is supported by an MSc scholarship from the mission department, Ministry of Higher Education (MoHE), Government of Egypt which is gratefully acknowledged.

\bibliographystyle{IEEEtran}
\bibliography{references.bib}

\begin{appendices}

\section{Derivation of Fast Node Count Estimator} \label{secCountEstimator}
\begin{figure}[h]
	\centering
	\includegraphics[width=\columnwidth]{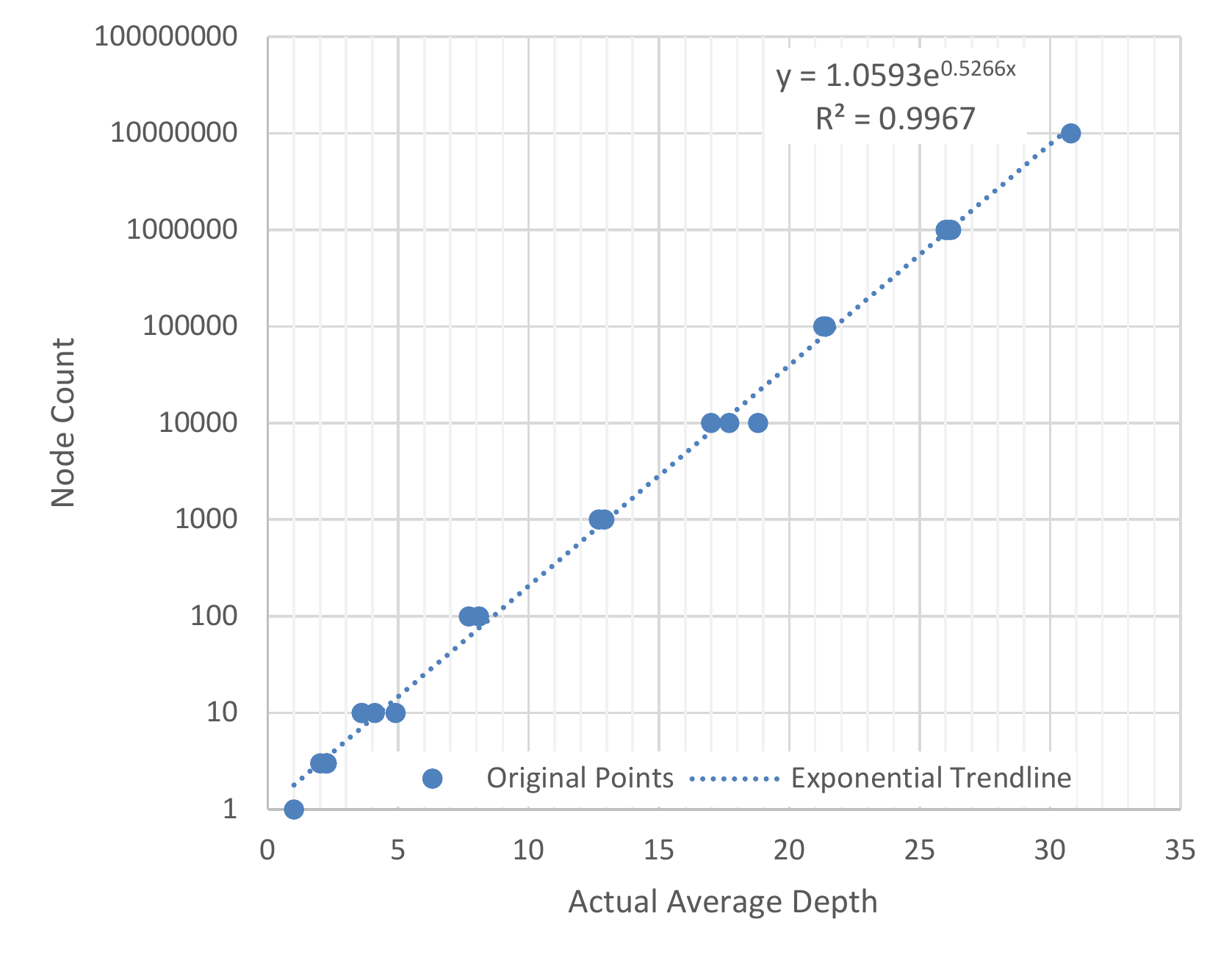}
	\caption{\label{figNodeCountEstimator}Actual Average Depth-Node Count Relation}
\end{figure}
The probing process is terminated based on a fast node count estimator based on tree depth. We derived its equation based on empirical data as shown in Fig.~\ref{figNodeCountEstimator}; for a variable number of random trees, the figure plots the actual average depth, $d$, against the corresponding actual number of nodes, $n$. We used least square exponential fitting to obtain the corresponding formula (\ref{eqNodeCountEstimator}):

\begin{equation}
n=1.0593\mathrm{e}^{0.5266d}
\label{eqNodeCountEstimator}
\end{equation}
The formula provides for $r^{2}\ge0.99$ that closely matches the relation.
\end{appendices}

\end{document}